\documentclass{iaus}
\usepackage{graphicx}
\title[ALFALFA: the Local HI Universe] 
{ALFALFA: HI Cosmology in the Local Universe}

\author[Giovanelli]   
{Riccardo Giovanelli}

\affiliation{Center for Radiophysics and Space Research,
Cornell University, Ithaca, NY 14853, \break email: riccardo@astro.cornell.edu}

\pubyear{2007}
\volume{244}  
\pagerange{xxx-xxx}
\date{?? and in revised form ??}
\jname{Dark Galaxies and Lost Baryons}
\editors{J.I.Davies \& M.J. Disney, eds.}

\def\kms{km~s$^{-1}$}
\def\etal{{\it et al.}}
\def\sqd{{deg$^{2}$}}
\def\arcmin{$^{\prime}$}
\def\arcsec{$^{\prime\prime}$}

\begin{document}

\maketitle

\begin{abstract}

For the last 25 years, the 21 cm line has been used productively to investigate 
the large--scale structure of the Universe, its peculiar velocity field and the 
measurement of cosmic parameters. In February 2005 a blind HI survey that will
cover 7074 square degrees of the high latitude sky was started at Arecibo, using 
the 7-beam feed L-band feed array (ALFA). Known as the Arecibo Legacy Fast ALFA 
(ALFALFA) Survey, the program is producing a census of HI-bearing objects over a 
cosmologically significant volume of the local Universe. With respect to previous blind HI 
surveys, ALFALFA offers an improvement of about one order of magnitude in
sensitivity, 4 times the angular resolution, 3 times the spectral resolution,
and 1.6 times the total bandwidth of HIPASS. ALFALFA can detect $7\times10^4 D^2$ 
$M_\odot$ of HI, where $D$ is the source distance in Mpc. As of mid 2007, 44\% of the 
survey observations and 15\% of the source extraction are completed. 
We discuss the status of the survey and present a few preliminary results,
in particular with  reference to the proposed ``dark galaxy'' VirgoHI21.
\end{abstract}

\firstsection 

\section{What is ALFALFA?}\label{sec:what}

Comprehensive wide angle surveys of the extragalactic HI sky became possible with 
the advent of multifeed front--end systems at L--band. The first such system with
spectroscopic capability
was installed on the 64~m Parkes telescope in Australia, and has produced the 
excellent results of the HIPASS survey (Barnes \etal ~2001). The 1990s upgrade 
of the Arecibo telescope, which replaced its line feeds with a Gregorian subreflector 
system, made it possible for that telescope to host feed arrays, as proposed by 
Kildal \etal ~(1993). Eventually a 7-beam radio ``camera'',  named ALFA 
(Arecibo L--band Feed Array), became operational at Arecibo,
enabling large--scale mapping projects with the great sensitivity of the 
305--m telescope. A diverse set of mapping projects are now underway at that
observatory, ranging from extragalactic HI line, to Galactic line and continuum, 
to pulsar searches. ALFALFA, the Arecibo Legacy Fast ALFA Survey, aims to map 
the peri-- and extragalactic HI emission over 7074 \sqd ~of the high galactic 
latitude sky. This will require 
a total of 4130 hours of telescope time. Exploiting the large collecting area of 
the Arecibo antenna and its relatively small beam size ($\sim 3.5$\arcmin), 
ALFALFA will be eight times more sensitive than HIPASS with $\sim$four times 
better angular resolution. The combination of sensitivity and angular resolution 
allows dramatically improved ability in determining the position of HI sources, 
a detail of paramount importance in the identification of source counterparts at 
other wavelengths. Furthermore, its spectral backend provides 3 times better 
spectral resolution (5.3 \kms ~at $z = 0$) and over 1.4 times more bandwidth. These
advantages, in combination with a simple observing technique designed
to yield excellent baseline characteristics, flux calibration and
HI signal verification, offer new opportunities to explore the 
extragalactic HI sky. A comparison of ALFALFA and other past and current HI
surveys is given in Table 1. Data taking for ALFALFA was initiated in 
February 2005, and, in the practical context of time allocation 
at a widely used, multidisciplinary national facility like Arecibo, 
completion of the full survey is projected to require 6 years.

{\bf The main science goals} to be addressed by ALFALFA are:
\begin{itemize}
\item The determination and environmental variance of the HI mass function,
especially at its faint end and its impact on the abundance of low mass halos
\item The global properties of HI selected galaxy samples
\item The large--scale structure characteristics of HI sources, their
impact on the ``void problem'' and metallicity issues
\item Provide a blind survey for HI tidal remnants and ``cold accretion''
\item A direct characterization of the HI Diameter function
\item Investigate the low HI column density environment of disks
\item Map and help elucidate the nature of HVCs around the Milky Way (and
beyond?)
\item The ALFALFA survey area includes $\sim 2000$ continuum sources with
fluxes sufficiently large to make useful measurements of HI optical
depth for DLA absorbers at low z
\item ALFALFA will double the number of known OH Megamasers at intermediate
redshifts
\end{itemize}
 
{\bf The choice of survey parameters} for ALFALFA is summarily justified by the
following consideration. The minimum integration time per beam in seconds 
$t_s$ necessary for ALFA to detect an HI source of HI mass $M_{HI}$, width 
$W_{kms}$, at a distance $D_{Mpc}$ is
\begin{equation}
t_s \simeq 0.25 \Bigl({M_{HI}\over 10^6 
M_\odot}\Bigr)^{-2} (D_{Mpc})^4 
\Bigl({W_{kms}\over 100}\Bigr)^\gamma, \label{eq:ts}
\end{equation}
where the exponent $\gamma\simeq 1$ for $W_{kms}<300$ \kms and increases to 
$\gamma\simeq 2$ for sources of larger width. This means that the depth
of the survey, i.e. the maximum distance at which a given HI mass can be
detected, increases only as $t_s^{1/4}$: longer integrations than strictly
necessary to achieve scientific goals lead to rapidly diminishing returns.
A corollary of this scaling law is the fact that, once $M_{HI}$ is
detectable at an astrophysically satisfactory distance, it is more 
advantageous to maximize the survey solid angle than to increase the
depth of the survey through longer dwell times. In passing, we note that
the $t_s$ required to detect a given $M_{HI}$ at a given distance decreases
as the 4th power of the telescope diameter: Arecibo offers a tremendous
advantage because of its huge primary reflector. A drift survey has many 
important advantages, especially at a telescope with properties which vary 
across the (AZ,ZA) space, such as the Arecibo dish. A double--drift survey 
offers all--important additional advantages in terms of RFI excision. The
effective integration time per beam area of a double--drift survey with ALFA
is of order of 40 sec. This translates to a minimum detectable HI mass
of $2\times 10^7$ $M_\odot$ at the distance of the Virgo cluster and
in a catalog of sources with a median $cz\sim 7800$ \kms. 

\begin{figure}[ht]
\centerline{
\scalebox{0.55}{%
\includegraphics{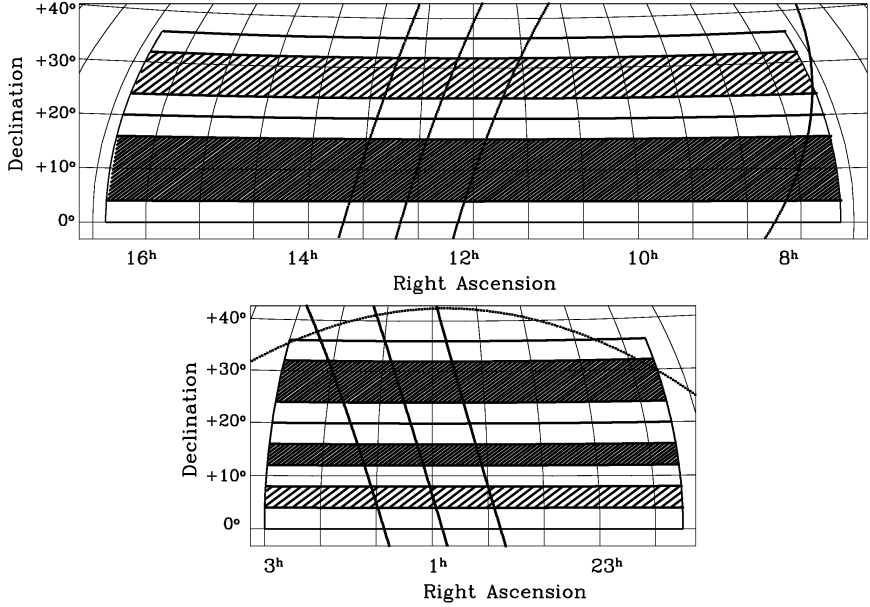}
}}
\vskip 1cm
\caption[]{\small
{Sky coverage of the ALFALFA survey, in the Virgo or Spring (upper) and anti-Virgo or Fall 
(lower) regions. In each panel, the thicker lines at constant RA or Dec. outline the 
boundaries of the survey area. The thick curves to the right of the upper panel and top of
the bottom panel mark b = $+20^\circ$ (upper) and $-20^\circ$ (lower) while the set of
three thick lines crossing each panel top to bottom trace SGL = $-10^\circ$, $0^\circ$ 
and $+10^\circ$. The regions in heavy shading correspond to parts
of the sky fully mapped by mid--2007, while light shading masks the regions with
started, but incomplete observations.
}}\label{skycov}
\end{figure}

{\bf The sky coverage of ALFALFA} includes the region from 0$^\circ$ to +36$^\circ$ 
in declination and from $22^h <$R.A.$< 3^h$ (the ``fall sky'') and
$7^h30^m <$R.A.$< 16^h30^m$ (the ``spring sky''), as illustrated in
Figure \ref{skycov}. Observations are carried out in drift mode, and each region 
of the sky is visited at two different epochs, separated by a few
months of Earth's orbital phase. No ``tracking L.O.'' is applied to
data taking. This fixed azimuth, ``minimum intrusion'' observing technique
delivers high data quality and extremely high observing efficiency. Because 
of its wide areal coverage and photometric accuracy, ALFALFA is providing a 
legacy dataset for the astronomical community at large, serving as the basis 
for numerous studies of the local extragalactic Universe. As of mid--2007, 44\% 
of the survey solid angle has been fully mapped. Further details on the design 
and progress of the survey can be seen in Giovanelli \etal (2005a), the
presentation by Martin ~in these proceedings and at the 
URL {\it http://egg.astro.cornell.edu/alfalfa}.

{\bf ALFALFA is an open collaboration}. Anybody with a legitimate scientific
interest and willing to participate in the development of the survey can
join. As of mid 2007, 68 individuals are participating in the ALFALFA
observations, processing or follow--up activities. Follow--up activities 
include: HI synthesis and single dish high sensitivity observations;
radio observations at other wavelength bands than HI; optical imaging in
broad band and H$_\alpha$ and for distance determination; optical spectroscopy 
for redshift confirmation, H$_\alpha$ rotation curve mapping and metallicity 
determination; infrared and UV imaging. Access to cataloged survey products
can be obtained at {\it http://arecibo.tc.cornell.edu/hiarchive/alfalfa}
and survey progress, guidelines for joining and other details can be obtained
at {\it http://egg.astro.cornell.edu/alfalfa}.

\begin{table}[!ht]
\caption{Comparison of Blind HI Surveys}
\smallskip
\begin{center}
\begin{tabular}{cccccccc}
\hline
\noalign{\smallskip}
Survey & Beam      & Area   & res    & rms$^a$ & V$_{med}$ & N$_{det}$ & Ref\\
       & (\arcmin) & (\sqd) & (\kms) &   & (\kms)    &           &    \\
\noalign{\smallskip}
\hline
\noalign{\smallskip}
AHISS   & 3.3 &    13 & 16 & 0.7 & 4800 &   65 &  $^b$ \\
ADBS    & 3.3 &   430 & 34 & 3.3 & 3300 &  265 &  $^c$ \\
WSRT    & 49. &  1800 & 17 & 18  & 4000 &  155 &  $^d$ \\
HIPASS  & 15. & 30000 & 18 & 13  & 2800 & 5000 &  $^{e,f}$ \\
HI-ZOA  & 15. &  1840 & 18 & 13  & 2800 &  110 &  $^g$ \\
HIDEEP  & 15. &    32 & 18 & 3.2 & 5000 &  129 &  $^h$\\
HIJASS  & 12. &  1115 & 18 & 13  & $^i$ &  222 &  $^i$\\
J-Virgo & 12. &    32 & 18 &  4  & 1900 &   31 &  $^j$\\
AGES    & 3.5 &   200 & 11 & 0.7 & 12000&      &  $^k$\\
ALFALFA & 3.5 &  7074 & 11 & 1.7 & 7800 &$>$25000& $^l$\\
\noalign{\smallskip}
\hline
\end{tabular}
{\small

$^a$ mJy per beam at 18 \kms ~resolution;
$^b$ Zwaan \etal ~(1997, \textit{ApJ} 490, 173);
$^c$ Rosenberg \& Schneider (2000, \textit{ApJSS} 130, 177);
$^d$ Braun \etal ~(2003, \textit{AAp} 406, 829);
$^e$ Meyer \etal ~(2004, \textit{MNRAS} 350, 1195);
$^f$ Wong \etal ~(2006, \textit{MNRAS} 371, 1855);
$^g$ Henning \etal ~(2000, \textit{AJ} 119, 2696);
$^h$ Minchin \etal ~(2003, \textit{MNRAS} 346, 787);
$^i$ Lang \etal ~(2003, \textit{MNRAS} 342, 738), HIJASS has a gap in velocity coverage between 4500-7500
\kms, caused by RFI;
$^j$ Davies \etal ~(2004, \textit{MNRAS} 349, 922);
$^k$ Minchin \etal ~(2007, \textit{IAU 233}, 227);
$^l$ Giovanelli \etal ~(2007, \textit{AJ} 133, 2569).
}
\end{center}
\end{table}

{\bf Data processing of ALFALFA} and signal extraction (Saintonge 2007) 
takes place through IDL applications developed at
Cornell University, fully embedded in the Virtual Observatory (VO)
environment. Cross--referencing of HI data with source catalogs and images
obtained at other wavelengths is enabled through all stages of data
processing. ALFALFA products can be accessed through VO--compatible software
tools: partial source catalogs are placed on the public domain as the survey
progresses. Precursor ALFA data (Giovanelli \etal ~2005b), previous archival 
pointed HI observations by the Cornell group for 9000 galaxies (this is 
already the largest collection of digital HI galaxy spectra in existence in 
the world) and the first ALFALFA catalog release (Giovanelli \etal ~2007) are 
available through a node of the Virtual Observatory at the Cornell Theory Center 
(see {\it http://arecibo.tc.cornell.edu/hiarchive}).
Team members have web access to preliminary, searchable source catalogs for the 
planning and execution of multiwavelength followup observations. These SQL searchable 
databases and plotting tools are made public as their associated
presentation papers are accepted for publication, as per VO requirement.
The data reduction, signal extraction and ancillary software has been exported 
to 15 sites where it is in regular use by team members. Well-developed documentation
and hands-on training in observation and reduction techniques are provided to new
members by ALFALFA experts at Cornell, to ensure commonality of standards.

\section{First Catalog Releases}\label{sec:firstcat}

Two catalogs of HI sources extracted from 3-D spectral data cubes have been 
submitted for publication in the first half of 2007, the first 
(Giovanelli \etal ~2007) for the Spring sky, the second (Saintonge \etal ~2007)
for the Fall sky. In the first catalog, now in the public domain, 
730 HI detections are cataloged and optical counterparts assigned, within the solid 
angle $11^h44^m <$ R.A.(J2000) $< 14^h00^m$ and $+12^\circ <$ Dec.(J2000) $<+16^\circ$
(which includes the northern part of the Virgo cluster), 
and redshift range $-1600$ \kms $~< cz < +18000$ \kms. In comparison, the HI Parkes All-Sky 
Survey (HIPASS) detected 40 HI sources in the same region, 2 of which are 
unconfirmed by ALFALFA. ALFALFA HI detections are reported for three
distinct classes of signals: (a) detections with signal-to-noise ratio
S/N $>$ 6.5; (b) high velocity clouds
in the Milky Way or its periphery; and (c) signals of lower S/N (to $\sim$ 4.5)
which coincide spatially with an optical object of known, matching redshift.

Although this region of the sky has been heavily surveyed by previous targeted
observations based on optical flux-- or size-- limited
samples, 69\% of the extracted sources are newly reported HI detections, an
indication of the fact that our collective criteria for selecting potentially
HI--rich targets have neglected most of the HI-rich population of objects.

\begin{figure}[ht]
\centerline{
\scalebox{0.8}{%
\includegraphics{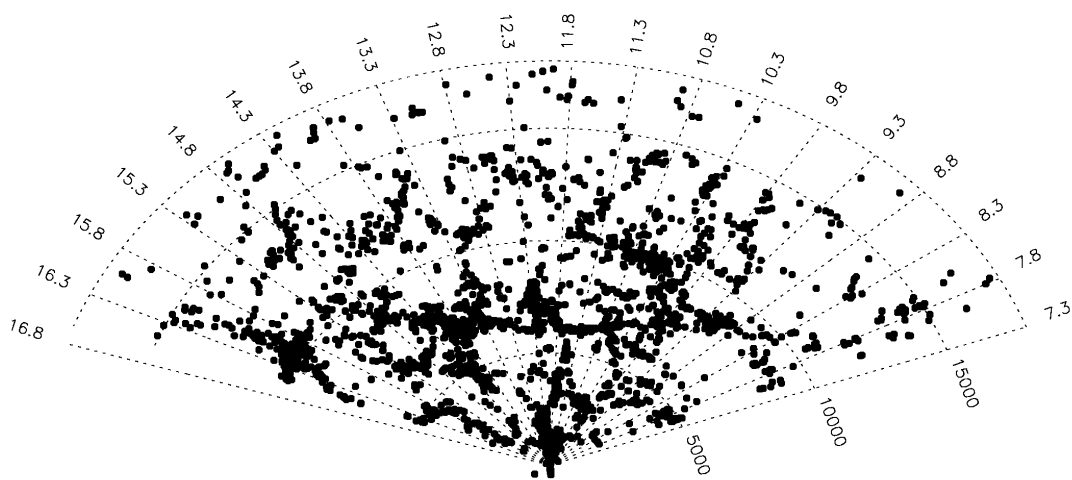}
}}
\vskip 1cm
\caption[]{\small
{Wedge plot of 2657 HI sources detected by ALFALFA in the region R.A.=[7.5$^h$--16.5$^h$],
Dec=[12$^\circ$--16$^\circ$], which represents 7.5\% of the survey. Note that due 
to RFI, ALFALFA is effectively blind in the redshift range 
between approximately 15000 and 16000 \kms. 
}}\label{cone}
\end{figure}

Signal extraction from ALFALFA data has been completed for some 15\% of the
survey area, as of mid--2007 and catalogs are in preparation for publication
for submission in the second half of 2007. The largest contiguous region
fully processed to date corresponds to a strip between 7.5$^h$ and 16.5$^h$ 
in R.A., 12$^\circ$ to 16$^\circ$ in Dec. A total of 2657 HI sources have
been detected in that region, amounting to about 7.5\% of the survey total
solid angle. The detection areal density of about 5 sources per sq. deg.
suggests that the full survey ``catch'' may eventually add to close to
30,000 HI sources, somewhat higher a number than estimated from survey
simulations. Figure \ref{cone} shows a wedge diagram of the strip data
set, neatly outlining the characteristics of the local Universe' s large
scale structure. Note that, due to the impact of local RFI, ALFALFA is
effectively blind in the redshift range $cz\sim$15000 to 16000 \kms.

\section{Statistical Characteristics of the Survey}\label{sec:stats}

\begin{figure}[ht]
\centerline{
\scalebox{0.13}{%
\includegraphics{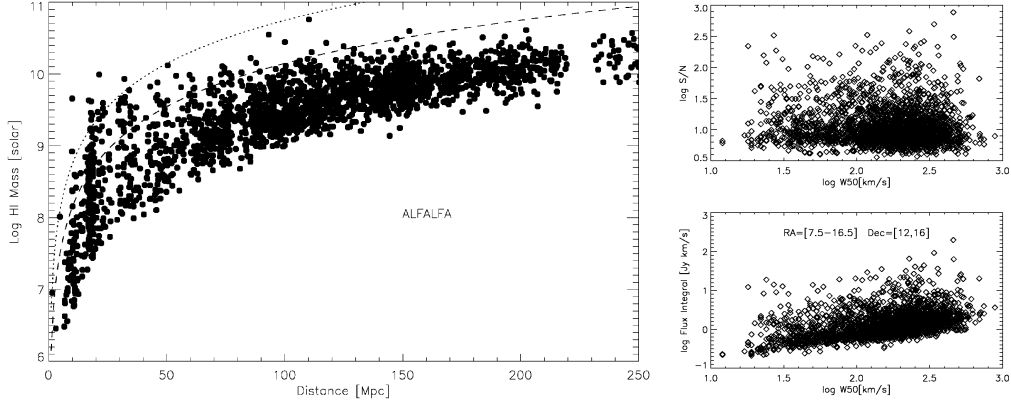}
}}
\vskip 1cm
\caption[]{\small
{{\bf Left:} Sp\" anhauer plot of 2657 HI sources detected by ALFALFA in the region R.A.=[7.5$^h$--16.5$^h$],
Dec=[12$^\circ$--16$^\circ$]. The two smooth lines identify respectively the completeness limit
(dotted) and the detection limit (dashed) for sources of 200 \kms ~linewidth for
the HIPASS survey. Note that due to rfi, ALFALFA is
effectively blind in the redshift range between approximately 15000 and 16000 \kms. 
{\bf Right:} Signal--to--noise ratio vs. velocity width (top) and Flux integral vs.
velocity width for the galaxies in the left--hand panel.
}}
\label{himd}
\end{figure}

Fig. \ref{himd} shows a HI mass vs. distance diagram of the HI sources in
Fig. \ref{cone}. Two smooth lines are overplotted, identifying respectively 
the completeness limit (dotted) and the detection limit (dashed) for sources 
of 200 \kms linewidth for the HIPASS survey. This diagram dramatically
illustrates the improvement ALFALFA represents, over previous surveys.
The median redshift of the catalog is $\sim$7800 \kms ~and its
distribution reflects the known local large scale structure. 
See Haynes' presentation in these proceedings for a discussion of the impact
of these observations on the faint end of the HI mass function.

For the same set of sources, top panel on the right of Fig. \ref{himd} displays 
S/N vs. velocity width, while the bottom panel displays the flux integral
vs. velocity width. The quality of the ALFALFA signal extraction is apparent: 
the S/N of detections exhibits no significant bias with respect to velocity width.
Spectroscopic HI surveys are not single flux limited. The flux limit is expected
to rise as $W^{1/2}$ for low velocity widths, changing to a linear rise for the
wider line profiles. Such a transition is observed near $\log W\simeq 2.5$. The
ALFALFA flux limit is $\sim 0.25$ Jy \kms ~for narrow lines, rising near 1 Jy \kms
~for the broadest ones.

The positional accuracy of HI sources is a very important survey parameter,
especially in the identification of HI sources with sources at
other wavelengths, as stressed by Disney (these proceedings). The quality of 
the positional centroiding of a source depends roughly linearly on source S/N
and inversely on the telescope beam angular size. 
Consider, for example, a source barely detected by HIPASS at S/N$\simeq 6.5$.
The error box of its positioning will have a radius of approximately 2.5'.
The same source can be detected by ALFALFA with $S/N\simeq 50$; as the Arecibo
beam is about 4 times smaller than that of Parkes, the positional error box
for the ALFALFA observation is $\sim 0.1'$, making an optical identification,
for example, far more reliable. Positional accuracy of ALFALFA sources 
averages 24\arcsec ~(20\arcsec ~median) for all sources with S/N $>$ 6.5 and is 
$\sim$17\arcsec ~(14\arcsec ~median) for signals S/N $>$ 12.

\section{Early Findings: Dark Galaxies?}\label{sec:early}

\begin{figure}[ht]
\centerline{
\scalebox{0.45}{%
\includegraphics{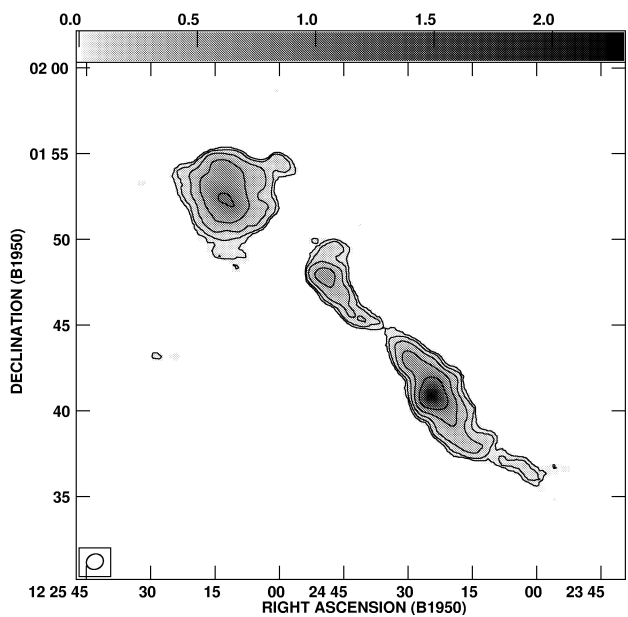}
}}
\vskip 1cm
\caption[]{\small
{Zeroth moment, HI VLA map of HI1225+01, as observed by Chengalur, Giovanelli
\& Haynes (1995) [thanks to Ekta and Jayaram for retrieving the image]. 
}}\label{HI1225}
\end{figure}

What should we refer to as a ``dark galaxy'' (perhaps a misnomer)? Within the
framework of this symposium, a dark galaxy would be a starless halo, yet
detectable at other than optical wavelengths, possibly in HI or through
lensing experiments. Such objects are likely to exist, but hard to find.
Within the CDM galaxy formation paradigm, such objects would have 
relatively low mass, were unable to form stars before re--ionization and
either lost their baryons or were prevented from cooling them thereafter,
by the IG ionizing flux (see presentations by Hoeft and by Yepes, these
proceedings). Yet we know of low mass galaxies in the Local Group which
not only made stars early on, presumably before re--ionization, but they 
were also capable of retaining cold gas and make stars at later cosmic times. 
Why then should we not expect the existence of low mass systems that were unable
to form stars but have retained baryons and have been able to cool them,
as the IG ionizing flux rarefies? We have extremely little observational
evidence for the existence of such systems. In my view, the best ``historical''
example for a ``dark galaxy''  is the SW component of the system
HI1225+01 discovered by Giovanelli \& Haynes (1989). Fig. \ref{HI1225} is a
zero-th moment VLA HI map of the system, by Chengalur, Giovanelli
\& Haynes (1995). The NE component has an optical counterpart
near its center. The SW component, however, does not: it exhibits evidence
of rotation $(V_{rot}\simeq 14$, a total mass near $10^9$ M$_\odot$ within
a 3' ($\sim 12$ kpc) radius and $M_{HI}/L>200$. However, it is not an 
isolated object and it cannot be excluded that it originated from a high
speed tidal encounter of the NE component with a now remote passer--by,
as the system lies in the outskirts of the Virgo cluster. The burden on
observers is that of finding isolated systems resembling HI1225+01SW.
See Kent's presentation (these proceedings) for possible candidates.
The bottom line is that, with the exception of systems found in the
periphery of the Virgo cluster, ALFALFA has not unambiguously discovered
any such system, so far.

\section{The Case of VirgoHI21}\label{sec:virgoHI21}

\begin{figure}[ht]
\centerline{
\scalebox{0.75}{%
\includegraphics{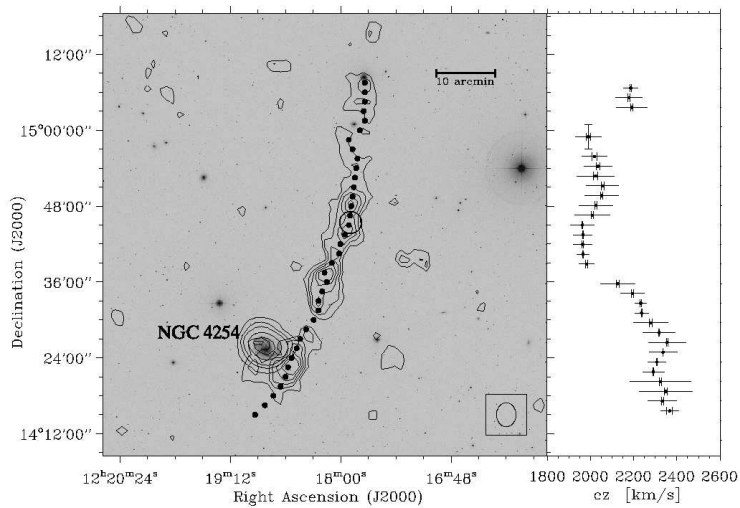}
}}
\vskip 1cm
\caption[]{\small
{Left: HI column density contours extracted from the ALFALFA
survey dataset, superposed on the SDSS image and centered on the position
of Virgo~HI21 (Minchin \etal ~2005a). Small
integer numbers superposed on the map indicate the locations of beam centers
of the Arecibo observations. Contours along the stream correspond to
0.35, 0.50, 0.70, 0.87, 1.0 Jy km s$^{-1}$ beam$^{-1}$, integrated from 1946 
to 2259 \kms; contours around N4254 correspond to 10, 15, 20, 30, 40 
Jy km s$^{-1}$ beam$^{-1}$, integrated from 2259 to 2621 \kms. Right: The velocity
of the HI emission peak as seen in each high S/N pointing. The horizontal bars
indicate the half--power full velocity width of the HI emission. VirgoHI21 was
identified as a section of the HI stream extending from $14^\circ 41'$ to $14^\circ 49'$.
}}\label{n4254}
\end{figure}

A candidate ``dark galaxy'' was discovered and dubbed ``VirgoHI21''
(Minchin \etal ~2007 and refs therein and these proceedings). The
discovery at Jodrell Bank was corroborated by Arecibo and WSRT observations.
VirgoHI21 lies some 100 kpc N of NGC4254, in the NW periphery of the Virgo 
cluster, projected $\sim 1$ Mpc from the cluster center. ALFALFA also
detects it, but the ALFALFA data strongly suggests a different interpretation
for the nature of VirgoHI21. Fig. \ref{n4254} (left) displays contours of HI 
flux of ALFALFA data, superimposed on an optical image, showing a gas
streamer extending some 250 kpc N of NGC 4254. The velocity field of the 
stream, which matches the velocity of NGC4254 to the S, is shown on the
right hand side panel of the figure. VirgoHI21 is the bright section of
the HI stream extending from  $14^\circ 41'$ to $14^\circ 49'$. The HI
mass in the disk of the galaxy is $4.3\times 10^9$ M$_\odot$ and that associated with 
the stream is $5.0\pm0.6\times 10^8$ M$_\odot$. One of the driving arguments
for Minchin \etal's interpretation of VirgoHI21 as an isolated disk
galaxy is the gradient seen in the velocity field; ALFALFA data shows
that gradient to be just a part of the varying, large--scale velocity field
along the stream.

NGC 4254 is a system well known for its prominent $m=1$ southern spiral
arm. It is reasonable to postulate that this special feature is related
with the existence of the stream. Note the following:
\begin{itemize}
\item NGC 4254 moves at a large velocity with respect  to the cluster
($>1000$ \kms) and lies at a projected distance of $\sim$ 1 Mpc from M87.
\item The prominent m=1 arm is visible in the gas and in the disk stellar
population: gravity, rather than hydro phenomena such as ram pressure, is
at work.
\item The HI mass in the stream is only $\sim 10$\% of that in the NGC 4254
disk: the disturbance of NGC 4254 is relatively mild (it would not, in fact
be classified as an HI deficient galaxy).
\item The velocity field of the stream shows the coupling of the tidal
force and the rotation of NGC 4254, which suggests an iinteresting
timing argument:
\begin{enumerate}
\item the stream exhibits memory of a full rotational cycle of the NGC 4254 disk;
\item from the NGC 4254 VLA map of Phookum, Vogel and Mundy (1993), we can get
the present outer radius of the HI disk (18.5 kpc) and the rotational
velocity at that radius (150 \kms); from those we compute a rotation
period of $\simeq 800$ Myr.
\end{enumerate}
\item Hence we estimate that the tidal encounter which gave rise to the
stream initiated some 800 Myr ago, a time comparable with the cluster
crossing time.
\end{itemize}
We conclude that the most reasonable interpretation of the system is that
of a relatively mild episode of harassment, resulting from the high
speed passage of NGC 4254 through the cluster periphery. These results are
discussed in greater detail in Haynes, Giovanelli and Kent (2007).

In these proceedings, P.A. Duc (see also Duc \& Bournaud 2007) presents the results 
of a simulation of a high speed encounter of NGC 4254 with another peripheral
cluster galaxy. The simulation matches extremely well both the morphology
and the velocity field of the stream. The culprit responsible for the
harassment of NGC 4254 could now be located
some 400 kpc to NW of NGC 4254. Duc \& Bournaud speculate that, given its location and
velocity, the culprit could be M 98 = NGC 4192. ALFALFA finds an extended
HI appendage apparently emanating from that galaxy, as is briefly discussed
in the next section.

The overall evidence for VirgoHI21 to be part of the phenomenology associated
with a tidal episode of harassment, rather than an isolated ``dark galaxy''
is thus quite strong.

\section{Other Virgo Cluster Streams: an HI Cemetery}\label{sec:cemetery}

As shown in the presentation by Kent, these proceedings, ALFALFA finds 
a number of HI features lacking an obvious optical counterpart. One 
important characteristic is in common with such systems found by ALFALFA: they appear
to preponderantly be found in the Virgo cluster region. In addition,
several of the systems share similarities with the NGC 4254 system:
HI streams associated with disturbed galaxies, extending a few to
several hundred kpc, suggesting that their origin is possibly related
to high speed encounters. In this section, a brief report is given
on two additional such systems.

{\bf The NGC 4532/DDO 137 pair} is located in the southern part of the Virgo
cluster. Using Arecibo, Hoffman \etal ~(1993) first discovered evidence 
that significant amounts of HI is present outside the galaxies' disks.
This result was confirmed by VLA observations of the system (Hoffman
\etal ~1999), showing numerous shreds of intergalactic HI spread over
tens of kpc from the optical galaxies. R. Koopmann (these proceedings) 
has recently discovered a complex HI stream system extending from the
galaxy pair to the S, over more than $1.5^\circ$ ($\sim 450$ kpc),
with numerous knots which are potential sites for the formation of tidal dwarfs.

{\bf NGC 4192 = M 98} is a galaxy located in the NW periphery of
the Virgo cluster, with a negative heliocentric velocity (-142 \kms)
indicative of large relative motion with respect to the cluster itself.
A number of HI clouds with velocities between 60 and 100 \kms ~are
found in its vicinity, spectrally well separated from Milky Way
emission. They would typically be cataloged as High (or Intermediate)
Velocity Clouds, perigalactic (or Local Group) dwellers, and
most of them may very well be just that. Some of the clouds, however,
appear to stream out of M 98, stretching over more than $1^\circ$,
well matched both spatially and kinematically to the disk of M 98. 
The characteristics of this system are under close scrutiny. Of
particular interest is the  fact that --- as mentioned in the preceding
section ---, Duc \& Bournaud suggest that M 98 may be the culprit
responsible for the harassment of NGC 4254, observed in the form of 
the stream VirgoHI21 is a part of.

\section{Conclusions}\label{sec:concl}

ALFALFA, the Arecibo Legacy Fast ALFA survey, is well underway. Designed
to map $\sim 7000$ sq. deg. of high galactic latitude sky in the HI line, 
observations for ALFALFA are 44\% complete
as of mid--2007. The first source catalog has been released and
others are in preparation. Survey products are being placed in
the public domain and accessible through NVO compatible software
tools. The fully processed data are delivering an average detection 
rate of $\sim 5$ sources deg$^{-2}$, with peaks 10--20 times
higher in high density regions such as group and cluster peripheries. 
The positional accuracy of the ALFALFA HI sources
is quite high, allowing unambiguous identification of $\sim 95$\% of them with 
optical counterparts. The median $cz$ is 7800 \kms, yielding sampling
of a fair cosmic volume.

Very few HI sources are not associated with optical counterparts. Those which
are not generally appear as parts of tidal streams or otherwise generated
appendages of known optical galaxies. The most dramatic examples of such
objects in the sets of ALFALFA data examined so far are found in the
vicinity of the Virgo cluster. Of particularly current interest is the
stream connected to NGC 4254, which icludes VirgoHI21. ALFALFA data
strongly favor the interpretation that VirgoHI21 is part of a tidal feature
that resulted from a high speed encounter, rather than a {\it bona fide} 
dark galaxy.

\begin{acknowledgments}
This work has been supported by NSF grants AST--0307661,
AST--0435697, AST--0607007. The Arecibo 
Observatory is part of the National Astronomy
which is operated by Cornell University under
a cooperative agreement with the National Science Foundation.
\end{acknowledgments}




\end{document}